\documentclass[aps,twocolumn,showpacs,prl,floatfix]{revtex4}
\usepackage{graphicx}
\usepackage{epsf}
\usepackage{wrapfig}
\usepackage{epsfig}
\def\bk{{\mbox{\boldmath$k$}}}
\def\bq{{\mbox{\boldmath$q$}}}

\def\bp{{\mbox{\boldmath$p$}}}

 \def\br{{\mbox{\boldmath$r$}}}

   \def\bp{{\mbox{\boldmath$p$}}}

\def\br{{\mbox{\boldmath$r$}}}
\def\b0{{\mbox{\boldmath$0$}}}

\def\bk{{\mbox{\boldmath$k$}}}
\def\bq{{\mbox{\boldmath$q$}}}

\def\bp{{\mbox{\boldmath$p$}}}
\def\boldkappa{{\mbox{\boldmath$\Delta$}}}

\def\br{{\mbox{\boldmath$r$}}}
\def\b0{{\mbox{\boldmath$0$}}}

\newcommand{\ra}{\,\rangle}
\newcommand{\la}{\,\langle}

\def \b #1{ {\bf #1}}
\newcommand{\be}{\begin{eqnarray}}
\newcommand{\ee}{\end{eqnarray}}

\def \b #1{ {\bf #1}}

\def \b #1{ {\bf #1}}

     \font\tenbifull=cmmib10 scaled 1200 
     \font\tenbimed=cmmib9
     \font\tenbismall=cmmib7
       \def\bmit{\fam9 }
\mathchardef\bbkappa="7114
\mathchardef\bbrho="711A
\mathchardef\bbsigma="711B
\mathchardef\bbtau="711C
\mathchardef\bbvarrho="7125
\mathchardef\bbvarsigma="7126
\mathchardef\bbxi="7118
\def\boldkappa{{\bmit\bbkappa}}
\def\boldrho{{\bmit\bbrho}}

\begin{document}
\vskip 2mm \date{\today}\vskip 2mm
\title{Non-factorized calculation of the
 process $\bf ^3He(e,e'p)pn$ at medium energies}
  \author{M. Alvioli}
  \address{104 Davey Lab, The Pennsylvania State University, University Park, PA 16803, USA}
\author{C. Ciofi degli Atti}
\author{L.P. Kaptari}
\altaffiliation{On leave from  Bogoliubov Lab.
      Theor. Phys.,141980, JINR,  Dubna, Russia}
\address{Department of Physics, University of Perugia and
      Istituto Nazionale di Fisica Nucleare, Sezione di Perugia,
      Via A. Pascoli, I-06123, Italy}
\begin{abstract}
The cross section and the transverse-longitudinal asymmetry
$A_{TL}$ of the three-body-break-up (3BBU) process $^3He(e,e'p)pn$
have been calculated by a non-factorized and parameter-free
approach based upon realistic few-body wave functions
corresponding to the $AV18$ interaction, treating the rescattering
 of the struck nucleon  within  a generalized eikonal approximation.
The results of  calculations exhibit a good agreement with recent
JLab experimental data  and show that  the Final State Interaction
(FSI)  of the struck proton in the region of missing momentum
 $300 \lesssim p_m \lesssim 600\,\, MeV/c$ and
  removal energy $E_m \gtrsim p_m^2/4m_N$ is dominated
  by single nucleon rescattering,  in agreement with nucleon-nucleon
correlation predictions.
\end{abstract}
\pacs{24.10.-i,25.10.+s,25.30.Dh,25.30.Fj} \maketitle

 In a
previous paper  \cite{OurPRLNonF} the cross section of the 2-body
break up (2BBU)  $^3He(e,e^\prime p)^2H$ recently measured at JLab
\cite{jlab1} in the missing momentum range  $ 0 \lesssim p_m
\lesssim 1000\,\, MeV/c$,  has been theoretically described by a
fully non-factorized and parameter-free theoretical approach in
which: i) initial state
  correlations (ISC) in the target nucleus $^3He$  have been taken care of  by using state-of-the-art
  few-body wave functions obtained ~\cite{pisa} by a variational solution of the Schr\"odinger
  equation containing realistic nucleon-nucleon interactions
   \cite{av18};
ii) FSI have been treated by a Generalized Eikonal Approximation
(GEA)~\cite{mark}, which represents an extended Glauber  approach
(GA) based upon
 the evaluation of the relevant
  Feynman diagrams,  which describe  the rescattering of the
   struck nucleon in the final state, in analogy with the
  Feynman  diagrammatic approach  developed for the treatment
  of elastic hadron-nucleus scattering  \cite{gribov,bertocchi}.
The results obtained in \cite{OurPRLNonF}, in agreement with other
types of realistic calculations \cite{Schiavilla}, have shown
that: i) the FSI of the struck proton in the 2BBU channel can
satisfactorily be described within single and double rescattering
of the struck proton (cf. Fig. 1 of
 \cite{OurPRLNonF}); ii) the low momentum part ($p_m \lesssim
300\,\, MeV/c$) of the cross section is practically not affected
by FSI and can even be described  within a Factorized
Approximation (FA),  which assumes a factorization of the
electron-Nucleus cross section into a product of the
electron-nucleon cross section and the nucleon Spectral Function;
on the contrary at $300 \lesssim p_m \lesssim 1000\,\, MeV/c$, FSI
becomes very important, particularly at "negative" values of the
missing
 momentum (when $\phi$,  the   angle between the scattering and
 reaction planes, equals zero), and  the  FA breaks down
  (cf. Fig. 2 of
 \cite{OurPRLNonF}); iii)  in  the region $300
\lesssim p_m \lesssim 700\,\, MeV/c$  the effects of FSI are
almost entirely exhausted by single rescattering (we will come to
this point later on) (cf. Fig. 2 of
 \cite{OurPRLNonF}); iv) the $A_{TL}$ asymmetry is extremely
sensitive to FSI and cannot be explained within a FA (cf. Fig. 3
of
 \cite{OurPRLNonF}). Following
 \cite{OurPRLNonF}, we have undertaken the calculation of the
 3-Body-Break-Up (3BBU) channel
 $^3He(e,e^\prime p)pn$ using the same approach of
\cite{OurPRLNonF}; our results will be compared with the
experimental data of Ref. \cite{benmo}.  By considering the
general case of a target nucleus $A$, the relevant quantities
which characterize the  process $A(e,e'p)(A-1)$
 are the energy and momentum transfer $\nu$ and
$Q^2$, respectively,
   the
missing momentum\,\, ${\b p}_m =\bq - \bp_1$  (i.e.   the momentum
of the recoiling system  $A-1$), and  the\,\, {\it missing energy}
{ $E_m= \sqrt{P_{A-1}^2}+m_N -M_A \,\, = \nu - T_{A-1} -
T_{\bp_1}=|E_{A}| - |E_{A-1}| + E_{A-1}^*=E_{min}+E_{A-1}^*$.
Here, $\bp_1$ is the momentum of the detected proton,  $E_A
(E_{A-1})$ the (negative) ground state energy of the target
(recoiling) nucleus and $E_{A-1}^*$ the intrinsic excitation
energy of the latter ($E_{A-1}^* = E_{m}-|E_{min}|$). The cross
section of the process has form
\begin{widetext}
\be
 \frac{d^6 \sigma}
  {d \Omega'   d E' ~ d \Omega_{p_1}   dE_{p_1}}  =
\left |\frac{{\bf p}_1^2}{ \frac{|{p}_1|}{E_{{p}_1}}+
 \frac{|{\bf p}_1|-|{\bf q}|\cos\theta}{E_{A-1}^*}}\right |\frac{dE_m}{dE_{{p_1}}}
  \sigma_{Mott} ~ \sum_i ~
V_i ~ W_{i}^A( \nu , Q^2, {\b p}_m, E_m)
 \label{cross}
 \ee
 \end{widetext}
 where $i \equiv\{L,T, TL, TT\}$,  $V_i$  are  kinematical factors
and  the nuclear structure functions $W_i^A$ result from proper
combinations of the polarization vector of the virtual photon,
$\varepsilon_{\lambda} ^{\mu}$,  and the hadronic tensor,
$W_{\mu\nu}^{A}$, the latter depending  upon the nuclear current
operators ${\hat J_\mu^A}(0)$. We consider the interaction of the
incoming virtual photon $\gamma^*$ with a bound nucleon (the
active nucleon) of low virtuality ($p^2\sim m_N^2$) in the
  quasi-elastic  kinematics i.e.  corresponding to $x\equiv Q^2/2m_N\nu\sim 1$. In
 quasi-elastic kinematics, the virtuality  of the struck nucleon after
    $\gamma^*$-absorption  is also rather low and,
    provided ${\b p}_1$  is sufficiently high, nucleon rescattering
    with the "spectator" $A-1$  can be   described to a large extent in terms of  multiple
    elastic scattering processes in the
    eikonal approximation. Let us now consider the process $^3He(e,e'p)pn$.  In co-ordinate representation the initial three-
    body wave function  is $\Phi_{1/2,{\cal M}_3}(\br_1,\br_2,\br_3)$, whereas the
wave function of the final state is
\begin{widetext}
 \be
\Psi_f^*={\hat{\cal A}} \left[ \chi_{T_f,\tau_f}\,
S_\Delta^{FSI}(\br_1,\br_2,\br_3)\,e^{-i {\bf p}_1 {\bf r}_1}\,
e^{-i {\bf P}_{m} (\br_2+\br_3)/2} \,\Phi^{ {\bf k}_{rel}*}_{S_f
{\sigma}_f}(\br_2-\br_3)\chi_{\frac{1}{2},\lambda_f}^+ \right]
\label{states}
 \ee
\end{widetext}
 where  ${\hat{\cal A}}$ is a proper antisymmetrization
operator,  ${\chi}_{T_f, \tau_f}$
 denotes the isospin wave function,
 $\Phi^{{\bf k}_{rel}*}_{S_f {
\sigma}_f}(\br_2-\br_3)$ describes the relative motion of two
interacting particles in the continuum and, eventually,
$\chi_{\frac{1}{2},\lambda_f}$ is the spin wave function of the
struck nucleon.  The quantity
\begin{widetext}
 \be
S_\Delta^{FSI}(\br_1,\br_2,\br_3)= 1 + T_{(1)}^{FSI}(\br_1,\br_2,\br_3)+
T_{(2)}^{FSI}(\br_1,\br_2,\br_3)
 \label{smatrix}
 \ee
 \end{widetext}
is the GEA  S-matrix, which describes both the case of no FSI, as well as  single and double rescattering
 of the struck nucleon off the  spectator nucleons \cite{nashPR,mark}.
 By introducing the corresponding Jacobi coordinates ${\bf R}$, $\boldrho$ and
 $\br$, the relevant matrix elements, can be computed solely in terms of relative
 coordinates  $\boldrho$ and
 $\br$.  For ease of presentation we will give  explicit expressions of   the
single rescattering contribution only, which has the form
${T}_{(1)}^{FSI}(\boldrho,
\br)=-\sum\limits_{i=2}^3\theta(z_i-z_1){\rm
e}^{i\Delta_z(z_i-z_1)} \Gamma (\b {b}_1-\b{b}_i)$, where $\br_i
\equiv(z_i,  \b b_i)$. It can be seen that, unlike the usual GA,
the GEA S-matrix gets also a contribution  from  a parallel
momentum $\Delta_z$  of pure nuclear origin   depending upon the
external kinematics and
 the removal energy of the struck proton (within the "frozen approximation"  $\Delta_z=0$,  and
 the usual Glauber profile is recovered).
By assuming that the nuclear current operator is a sum of
nucleonic currents $\hat j_\mu(i)$, its momentum space
representation resulting from  Feynman diagrams can be written as
follows
\begin{widetext}
\begin{eqnarray}&&
 J_\mu^{3} =\sum_\lambda
 \int\frac{d {\bf p}}{(2\pi)^3} \frac{d\boldkappa}{(2\pi)^3}
S^{FSI}_\Delta (\bp,\boldkappa) \la
\lambda_f|j_\mu(\boldkappa-\bp_m;\bq)|\lambda\ra {\cal O}
(\bp_m-\boldkappa,\bp,{\bf k}_{rel}; {\cal
M}_3,S_f,{\sigma}_f,\lambda) =\nonumber\\ &&
=J_\mu^{3(PWIA)}+J_\mu^{3(1)}+J_\mu^{3(2)}, \label{ja}
\end{eqnarray}
\end{widetext}
where $S_\Delta^{FSI}(\bp,\boldkappa) =\int d\br d\boldrho e^{-i\bp \br}
e^{i\boldkappa \boldrho} S_{\Delta}^{FSI}(\boldrho,\br)$, is the Fourier transform of the GEA
S-matrix (Eq. (\ref{smatrix})),
   ${\cal O}$ the nuclear overlap in momentum space
\begin{widetext}
\be \label{overl} {\cal O} (\bp_m -\boldkappa, \bp,{\bf k}_{rel}
\,; \,{\cal M}_3,S_f,{\sigma}_f,\lambda)= \int d\boldrho d\br
e^{i(\bp_m - \,\boldkappa)\, \boldrho} e^{i \bp\,
\br}\Phi_{\frac{1}{2},{\cal M}_3}(\boldrho,\br)\Phi_{S_f,{
\sigma}_f}^{{\bf k}_{rel}*}(\br)\chi_{\frac12 \lambda}^+ \ee
\end{widetext}
and, eventually,  $\la
\lambda_f|j_\mu(\boldkappa-\bp_m;\bq)|\lambda\ra$ is the nucleon
current operator, with
 $\lambda$ and $\lambda_f$ being the spin projections of the struck
proton before and after $\gamma^*$ absorption. Note, that in eq.
(\ref{overl}) the quantity $\boldkappa-\bp_m  \equiv{\bf k}_1$
represents exactly the momentum of the struck proton before the
electromagnetic interaction, with  $\boldkappa$ being the momentum
transfer in single  rescattering. In absence of any FSI
 the momentum space S-matrix is $S_\Delta^{FSI}(\boldkappa ,
{\bf p})=(2\pi)^6\delta(\boldkappa)\delta({\bf p})$ and only the PWIA contribution
 $J_\mu^{3(PWIA)}$  survives in Eq. (\ref{ja}).    When FSI is active,
 the contributions from single and double rescattering have to be taken into account; let
  us consider the former:  it results  from the single scattering momentum space term of
   $S_\Delta^{FSI}(\bp,\boldkappa)$, which has the following form
 \begin{widetext}
 \be
T_{(1)}^{FSI}(\bp,\boldkappa)=  \frac{(2\pi)^4}{k^*}
\frac{f_{NN}(\boldkappa_\perp)}{\boldkappa_{||}+\Delta_z-i\varepsilon}\left[\delta\left(
\bp-\frac{\boldkappa}{2}\right)+\delta\left(\bp+\frac{\boldkappa}{2}\right)\right]
\label{T1}
 \ee
 where $f_{NN}(\boldkappa_\perp)$  (the Fourier transform of the profile function
 $\Gamma (\b b)$)
represents the
  elastic scattering amplitude of two nucleons with  center-of-mass
momentum $k^*$. Placing Eq. (\ref{T1}) in Eq. (\ref{ja})  the
single scattering contribution $J_\mu^{3(1)}$ to the nuclear
current is obtained as follows
\be
&& \!\!\!\!\!\!\!\!\!\!\!\!\!\!\!\!
J_\mu^{3(1)}=\sum_\lambda  \int \frac{d\boldkappa}{(2\pi)^2 k^* }
\la \lambda_f|j_\mu(\bk_1;\bq)|\lambda\ra
\frac{f_{NN}(\boldkappa_\perp)}{\boldkappa_{||} + \Delta_{||}-i\varepsilon}
 \times\nonumber\\&&
\left[ {\cal O} (-\bk_1,\frac{{\boldkappa}}{2}; {\cal
M}_3,S_f,{\cal \sigma}_f,\lambda) +{\cal O} (-\bk_1
,-\frac{{\boldkappa}}{2}; {\cal M}_3,S_f,{ \sigma}_f,\lambda)
\right]; \label{single} \ee
\end{widetext}
 where   $\bk_1$, as already mentioned,  is the momentum of the
proton before $\gamma^*$ absorption and   $\boldkappa =\bk_1+ \bq
-\bp_1=\bk_1+\bp_m$  is  the momentum transfer in the $NN$
rescattering. The longitudinal part of the nucleon propagators has
been computed using the relation $[{\boldkappa_{||}+\Delta_z \pm
i\varepsilon}]^{-1}=\mp i\pi\delta(\boldkappa_{||}+\Delta_z) +
PV[{\boldkappa_{||} +\Delta_z}]^{-1}$.
 Note  that in the eikonal approximation  the trajectory of
 the fast  nucleon is a straight line so that all
 "longitudinal" and "perpendicular" components are defined
 with respect  to this trajectory, which means that  the $z$ axis
 has  to be directed along the  momentum of the detected fast proton.
  Looking at the structure of Eq. (\ref{single}), it can be seen that, due to the coupling of
the nucleonic current operator $\la
\lambda_f|j_\mu(\bk_1;\bq)|\lambda\ra$ with the nuclear overlap
integral, a factorized form for Eq. (\ref{single})  cannot be
obtained; the same holds for the double-scattering contribution
and, consequently, for the cross section.
 However, as shown in Ref.
\cite{nashPR}, if  the longitudinal part of the nucleonic current
can be disregarded, the factorization form can approximately be
recovered. Using the above formalism, and including the contribution from double rescattering $J_\mu^{3(2)}$,
the cross section (Eq.
(\ref{cross})) and the left-right asymmetry defined by
\be A_{TL}
=\frac{d\sigma(\phi=0^o)-d\sigma(\phi=180^o)}{d\sigma(\phi=0^o)+d\sigma(\phi=180^o)}.
\label{atl}
\ee have been calculated. Following  de Forest's prescription
\cite{forest}, the "CC1" form of the nucleon current operator has
been adopted; the elastic amplitude $f_{NN}$ has been chosen in
the  usual  form $f_{NN}(\boldkappa_\perp)=k^*
\frac{\sigma^{tot}(i+\alpha)}{4\pi}e^{-b^2\boldkappa_\perp^2/2}$,
where the slope parameter $b$, the total nucleon-nucleon cross
section $\sigma^{tot}$ and the ratio $\alpha$ of the real to the
imaginary parts  of the forward scattering amplitude, were taken
from world's experimental data. The results of our calculations
are shown
  in  Figs. \ref{Fig1}, \ref{Fig2} and \ref{Fig3}. In Fig. \ref{Fig1}
  the full non-factorized cross section at two values of $p_{m}$
   is compared with the PWIA result and with the
    experimental data; in this Figure the role played by single and
     double rescattering is clearly
    illustrated.  In  Fig.\ref{Fig2} the factorized and
    non-factorized results are compared at $p_m = 440\,\, MeV/c$  (similar results
    are obtained at higher
    values of $p_m$). Finally in Fig. \ref{Fig3}
the asymmetry $A_{TL}$ is exhibited.
\begin{figure}[!hpt]
\hskip -3mm
\includegraphics[width=.47\textwidth]{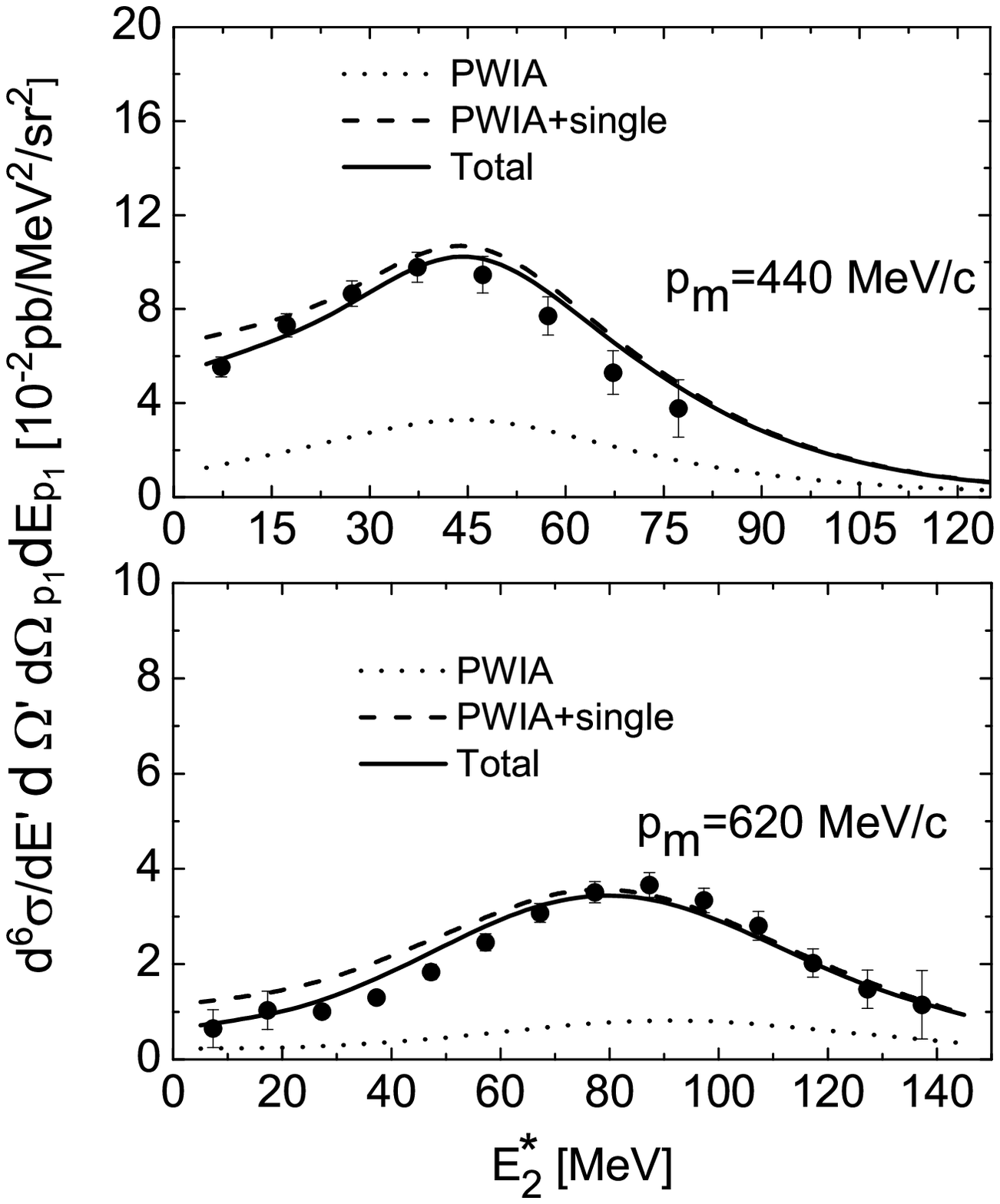} 
$A_{TL}$ of the three-body-break-up (3BBU) process $^3He(e,e'p)pn$
have been calculated by a non-factorized and parameter-free
approach based upon realistic few-body wave functions
corresponding to the $AV18$ interaction, treating the rescattering
 of the struck nucleon  within  a generalized eikonal approximation.
The results of  calculations exhibit a good agreement with recent
JLab experimental data  and show that  the Final State Interaction
(FSI)  of the struck proton in the region of missing momentum
 $300 \lesssim p_m \lesssim 600\,\, MeV/c$ and
  removal energy $E_m \gtrsim p_m^2/4m_N$ is dominated
  by single nucleon rescattering,  in agreement with nucleon-nucleon
correlation predictions.
\vskip -1mm\caption{The differential cross section (Eq.
(\ref{cross})) for the process $^3He(e,e^\prime p)pn$  calculated
at two values of the missing momentum  {\it vs.} the excitation
energy of the two-body final state $E_2^*= E_m-E_{min}$ ($E_{min}=
2m_p+m_n-M_3$). Dotted lines: PWIA approximation; dashed and solid
lines: non-factorized calculations  with single and double
rescattering in the final state, respectively. Experimental data
from \cite{benmo}.} \label{Fig1} \vskip -0.1cm
\end{figure}
   As in  Ref. \cite{OurPRLNonF},   no approximations have  been made in the evaluation
   of the single and double rescattering contributions:
   intrinsic coordinates have been used and the energy dependence of the profile
   function  has been taken into account in the properly chosen CM system of the
   interacting pair. Note that the experimental kinematics has been intentionally chosen so
   as to
   cover the theoretically predicted two nucleon correlation region characterized by
   a peak position given by
$E_{2}^*=E_m-E_{min}\sim 2m_N
\{\sqrt{(1/2)\,[1+\sqrt{1+(p_m/m_N)^2}]} -1 \}$ $\rightarrow
{p_{m}^2}/{4m_N}$ in  the non relativistic limit. It can be seen
that the location of the experimental peaks agrees with such a
prediction. The amplitude of the peak is largely affected by  FSI,
whose inclusion  brings the theoretical calculation in very good
agreement with the experimental data.
 The relevant effects of FSI can be explained by the fact
 \begin{figure}[!hpt]
\includegraphics[width=.47\textwidth]{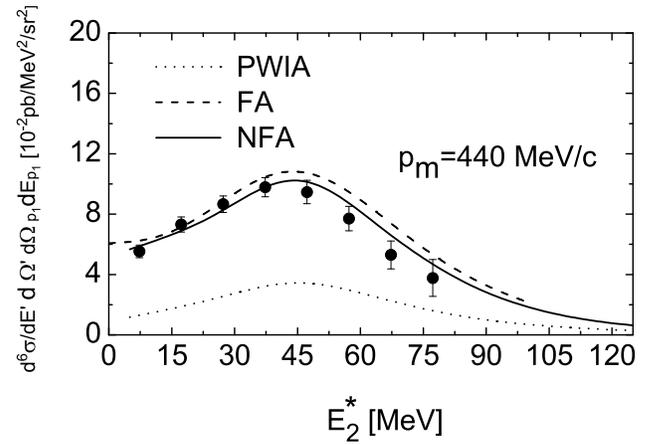} %
\vskip -1mm\caption{Comparison at $p_m=440 \,MeV$ of the results
of calculations performed within the non-factorized approach (NFA)
and the factorization approximation (FA).} \label{Fig2} \vskip
-0.1cm
\end{figure}
 that the  kinematics of the experiment is such that  the missing momentum   is always
 almost perpendicular to the momentum of the final proton and
   FSI is maximized. Concerning the effects of the
  latter, an interesting observation is in order: it can be seen that in the region of
  the correlation
  peak, and at higher values of $E_{2}^*$,  the effects from double rescattering is very tiny;
  the same
  appears to be the case in the 2BBU channel at
  $300 \lesssim p_m \lesssim 700\,\, MeV/c$. We consider this as another evidence that in
   these
  regions of missing momentum and energy FSI  mainly occurs within
  the correlated pair. In Fig. \ref{Fig2} the factorized and
  non-factorized results are compared at $p_m=440\,\,MeV/c$; it can be seen that
  the the non-factorized calculation differs by 5-7\% from the
  factorized one and better agrees with the experimental data.
  Eventually, in Fig. \ref{Fig3}  the
  transverse-longitudinal
  asymmetry  $A_{TL}$ (Eq. (\ref{atl})), calculated within the PWIA and taking into account
  FSI, is presented. It is well known that when  the explicit expressions of $V_i$
  and $W_i^A$ are placed
  in Eq.
  (\ref{atl}), the   numerator  is proportional to the transverse-longitudinal
  response $W_{TL}$, whereas the
 denominator does not contain $W_{TL}$ at all, which means that
 $A_{TL}$ is a measure of   the relevance of the transverse-longitudinal
 response relative to the other responses. In the
 $eN$ cross section the behavior of the asymmetry $A_{TL}$ is known
  to be a negative and  decreasing function of the missing momentum
  ~\cite{forest} and the same behavior should be expected in  $eA$ scattering
  within the PWIA.
\begin{figure}[!hpt]
\includegraphics[width=.47\textwidth]{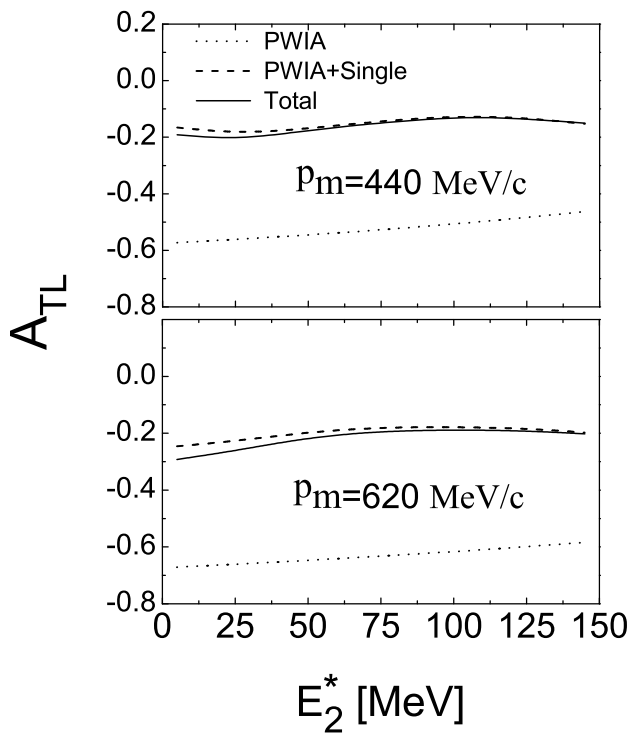}
\vskip -0.3cm\caption{The $A_{TL}$ asymmetry (Eq. (\ref{atl}))
calculated in correspondence of two values of $p_m$ {\it vs.} the
excitation energy of the two-body final state.} \label{Fig3}
\vskip -0.1cm
\end{figure}
The asymmetry  presented in Fig. \ref{Fig3} clearly
 shows that a measurement of the 3BBU channel at $\phi=0^o$
 would provide results not very different from the ones obtained
 at $\phi = 180^o$.
 To sum up, we have calculated in momentum space the cross section of the
 processes  $^3He(e,e'p)pn$, using realistic ground-state two- and three-body
 wave functions,  treating
 the FSI of the struck nucleon with the spectator nucleon pair
 within the  generalized eikonal  approximation.
 As in the case of the 2BBU calculation \cite{OurPRLNonF}, our
 approach is a parameter free  one, for it
only requires the knowledge of
 the nuclear wave functions, the FSI factor being fixed directly
 by  NN scattering data. At the same time, calculations are very involved
 mainly because of the complex structure of the wave function of Ref.~\cite{pisa},
which has to be firstly transformed to  momentum space and then
used in the calculations of multidimensional integrals, including
also the computation of principal values. The main results of our
calculations can be summarized as follows: i) a good agreement
between theoretical calculations and experimental data on the 3BBU
process $^3He(e,e'p)pn$ has been achieved; ii) FSI effects are
very relevant, but in the region of the correlation peak
($E_{2}^*\simeq\frac{p_{m}^2}{4m}$) and at higher energies
($E_{2}^*\ge \frac{p_{m}^2}{4m}$), they  can  be accounted for by
single rescattering, the double rescattering contribution being
negligible; this is  an implicit confirmation of the two-nucleon
correlation  picture, according to which the uncorrelated nucleon
(the A-2 system) is far away from the correlated pair and
practically does not contribute to the short-range rescattering
processes in the final state; iii) factorized and non-factorized
results may differ at
$\phi=180^o$ by at most 10\%.\\

The authors are  indebted to  the Pisa Group   for making
available the variational three-body  wave functions. Thanks are
due  to S. Gilad, M. Sargsian, S. Scopetta, and   M. Strikman  for
 useful  discussions.
L.P.K. is   indebted to  the Italian Ministry of University and
Research (MIUR) for a grant within the program "Rientro dei
Cervelli". He thanks the University of Perugia and INFN, Sezione
di Perugia, for  warm hospitality. M.A.  is supported by DOE grant
under contract DE-FG02-93ER40771

\end{document}